\documentclass[pdftex,letterpaper,10pt]{article}

\pdfoutput=1
\usepackage[dvips]{graphicx}
\usepackage{mathptmx}
\usepackage{amsmath}
\usepackage{latexsym}
\usepackage{amssymb}
\usepackage{abstract}
\usepackage{enumerate}
\usepackage[dvips]{color}
\usepackage{titlesec}
\usepackage{abstract}
\usepackage[square,sort&compress]{natbib}
\usepackage{ctable}
\usepackage[top=1.5in, bottom=1.5in, left=1.5in, right=1.5in]{geometry}
\newcommand{\footnoteremember}[2]{\footnote{#2}\newcounter{#1}\setcounter{#1}{\value{footnote}}}
\newcommand{\footnoterecall}[1]{\footnotemark[\value{#1}]}

\begin{document}

\title{Shocklets, SLAMS, and field-aligned ion beams in the terrestrial foreshock}

\author{L.B. Wilson III\footnoteremember{1}{NASA Goddard Space Flight Center, Greenbelt, Maryland, USA.}, A. Koval\footnoteremember{5}{Goddard Planetary Heliophysics Institute, University of Maryland Baltimore County, Baltimore, Maryland, USA.}\footnoterecall{1}, D.G. Sibeck\footnoterecall{1}, A. Szabo\footnoterecall{1}, C.A. Cattell\footnoteremember{2}{School of Physics and Astronomy, University of Minnesota, Minneapolis, Minnesota, USA.}, J.C. Kasper\footnoteremember{3}{Harvard-Smithsonian Center for Astrophysics, Harvard University, Cambridge, Massachusetts, USA.},\\ B.A. Maruca\footnoterecall{3}, M. Pulupa\footnoteremember{4}{Space Sciences Lab, University of California at Berkeley, Berkeley, California, USA.}, C.S. Salem\footnoterecall{4}, and M. Wilber\footnoterecall{4}}
\maketitle

\begin{abstract}
  We present Wind spacecraft observations of ion distributions showing field-aligned beams (FABs) and large-amplitude magnetic fluctuations composed of a series of shocklets and short large-amplitude magnetic structures (SLAMS).  We show that the SLAMS are acting like a local quasi-perpendicular shock reflecting ions to produce the FABs.  Previous FAB observations reported the source as the quasi-perpendicular bow shock.  The SLAMS exhibit a foot-like magnetic enhancement with a leading magnetosonic whistler train, consistent with previous observations.  The FABs are found to have T${\scriptstyle_{b}}$ $\sim$ 80-850 eV, V${\scriptstyle_{b}}$/V${\scriptstyle_{sw}}$ $\sim$ 1-2, T${\scriptstyle_{\perp, b}}$/T${\scriptstyle_{\parallel, b}}$ $\sim$ 1-10, and n${\scriptstyle_{b}}$/n${\scriptstyle_{i}}$ $\sim$ 0.2-14$\%$.  Strong ion and electron heating are observed within the series of shocklets and SLAMS increasing by factors $\gtrsim$5 and $\gtrsim$3, respectively.  Both the core and halo electron components show strong perpendicular heating inside the feature.
\end{abstract}


\section{Introduction}  \label{sec:introduction}
\indent  Collisionless shock waves are a ubiquitous phenomena in space plasmas and are thought to mediate and possibly produce some of the highest energy cosmic rays in the universe \citep{blandford87}.  They are also efficient accelerators of much lower energy particles \citep[\textit{e.g.}][]{kis07a} producing a region upstream of the shock transition called a foreshock.  The terrestrial foreshock has been examined in detail \citep[\textit{e.g.}][\textit{and references therein}]{eastwood05c}.  In the process of studying this region, five ion populations have been identified:  (1) field-aligned beams (FABs), (2) intermediate ions, (3) diffuse ions, (4) gyrating ions, and (5) gyrophase-bunched ions \citep[\textit{e.g.}][]{kis07a}.  All the ion species have been observed to be spatially well separated \citep{meziane11b}.  \\
\indent  Early observations \citep{greenstadt76a} suggested that FABs had their origin on field lines connected to a quasi-perpendicular (shock normal angle, or $\theta{\scriptstyle_{Bn}}$, $>$45$^{\circ}$) portion of the bow shock.  FABs are primarily composed of protons streaming along the ambient magnetic field away from the bow shock with temperatures (T${\scriptstyle_{b}}$) $\sim$80-600 eV, densities (n${\scriptstyle_{b}}$) 1-10$\%$ of the ambient solar wind density (n${\scriptstyle_{o}}$), and beam speeds up to 5 times the solar wind speed (V${\scriptstyle_{sw}}$) \citep{bonifazi81a, bonifazi81b, paschmann81a}.  FABs often have strong temperature anisotropies with T${\scriptstyle_{\perp, b}}$/T${\scriptstyle_{\parallel, b}}$ $\gtrsim$ 4-9 \citep{paschmann81a}.  They are typically observed in the absence of or near large magnetic fluctuations, but not simultaneously with these waves \citep{kis07a, meziane11b}.  \\
\indent  The terrestrial foreshock on the quasi-parallel side has a broad spectrum of large amplitude waves.  This spectrum consists of transverse Alfv\'{e}nic waves, right-hand polarized (in plasma frame) ultra-low frequency (ULF) waves near the ion gyrofrequency (f${\scriptstyle_{ci}}$), compressional magnetosonic waves \citep{hoppe83a}, magnetosonic-whistler mode waves \citep{hoppe81a}, and an ensemble of higher frequency (f $>$ 5-10 Hz) waves up to the electron plasma frequency (f${\scriptstyle_{pe}}$) \citep[\textit{e.g.}][\textit{and references therein}]{briand09a}.  We will focus on two specific types of waves in the foreshock, shocklets \citep{hoppe81a} and short large-amplitude magnetic structures (SLAMS) \citep{schwartz92a}.  Both types are thought to grow out of the ULF wave field, due to an interaction with gradients in the diffuse ion densities \citep{scholer03b}.  Both can radiate a dispersive higher frequency electromagnetic whistler precursor due to wave steepening and are always observed simultaneously with diffuse ion distributions \citep[\textit{e.g.}][\textit{and references therein}]{wilsoniii09a}.  Though SLAMS and shocklets have different scale sizes \citep{lucek02a}, they are more easily differentiated by their relative amplitudes.  Shocklets exhibit weak magnetic compression with $\delta$B/B${\scriptstyle_{o}}$ $\lesssim$ 2, while SLAMS are much stronger with $\delta$B/B${\scriptstyle_{o}}$ $>$ 2 and often exceeding a factor of 4.  \\
\indent  Previous studies have found FABs near SLAMS, but instrumental limitations prevented an examination of the evolution of the ion distributions across the SLAMS \citep{schwartz92a, wilkinson93a}.  More importantly, these studies showed indirect evidence that the SLAMS, not the bow shock, were producing the FABs locally.  \\
\indent  In this paper we report the first high time resolution observations of the evolution of FABs through large amplitude magnetic fluctuations, identified as shocklets and SLAMS, in the terrestrial foreshock.  The ion beams are more intense on the upstream (sunward) side of the SLAMS, suggesting a local source.  The paper is organized as follows:  Section \ref{sec:data} discusses the data sets and analysis techniques, Section \ref{subsec:foreobservations} provides an overview of relevant parameters, Section \ref{subsec:particledists} discusses the particle distribution observations, and Section \ref{sec:conclusions} presents our conclusions.
\section{Data Sets and Analysis}  \label{sec:data}
\indent  The magnetic field was obtained from the Wind dual, triaxial fluxgate magnetometers \citep{lepping95} sampled at $\sim$11 samples/s.  Full 4$\pi$ steradian low energy ($<$30 keV) ion and electron distributions were obtained from the Wind/3DP EESA and PESA particle detectors \citep{lin95a}.  For more details about the analysis of data from the 3DP instrument, see \citet{wilsoniii10a}.  The solar wind velocity (\textbf{V}${\scriptstyle_{sw}}$), average ion thermal speed (V${\scriptstyle_{Ti}}$), and average ion temperature (T${\scriptstyle_{i}}$) were determined with the 3DP PESA Low and SWE Faraday Cups (FCs) \citep{ogilvie95}.  Absolute electron densities were determined from the plasma line in the WAVES thermal noise receiver (TNR) instrument \cite{bougeret95a} and used as an estimate of the solar wind ion density (n${\scriptstyle_{i}}$).  To analyze the beams, we fit the FABs to bi-Maxwellians to determine estimates of the density (n${\scriptstyle_{b}}$), the temperature (T${\scriptstyle_{b}}$), and beam speed (V${\scriptstyle_{b}}$).  We will use these values to compare to previous observations and provide better constraints for theory.  \\
\indent  The wave vector, \textbf{k}, and the polarization with respect to the quasi-static magnetic field were determined using Minimum Variance Analysis (MVA) \citep{khrabrov98}.  The details of this technique are discussed in \citet{wilsoniii09a}.  We calculated the angles between the wave vector and the local magnetic field ($\theta{\scriptstyle_{kB}}$) and the solar wind velocity ($\theta{\scriptstyle_{kV}}$).  \\
\indent  We used two methods to determine the bow shock normal vector.  The first involved the use of the Rankine-Hugoniot conservation relations \citet{koval08a} with the parameters observed at the last crossing of the bow shock.  From this we derive a single shock normal vector.  The second method involved projecting the local magnetic field vector onto the surface of a model bow shock \citep{slavin81a}.  Once we determined the shock normal vector, we were able to calculate the shock normal angle, $\theta{\scriptstyle_{Bn}}$.
\section{Observations}  \label{sec:observations}
\subsection{Foreshock Observation Overview}  \label{subsec:foreobservations}
\begin{figure}[htb]
  \vspace{-5pt}
  \begin{center}
   {\includegraphics[trim = 0mm 0mm 0mm 0mm, clip, width=16cm]{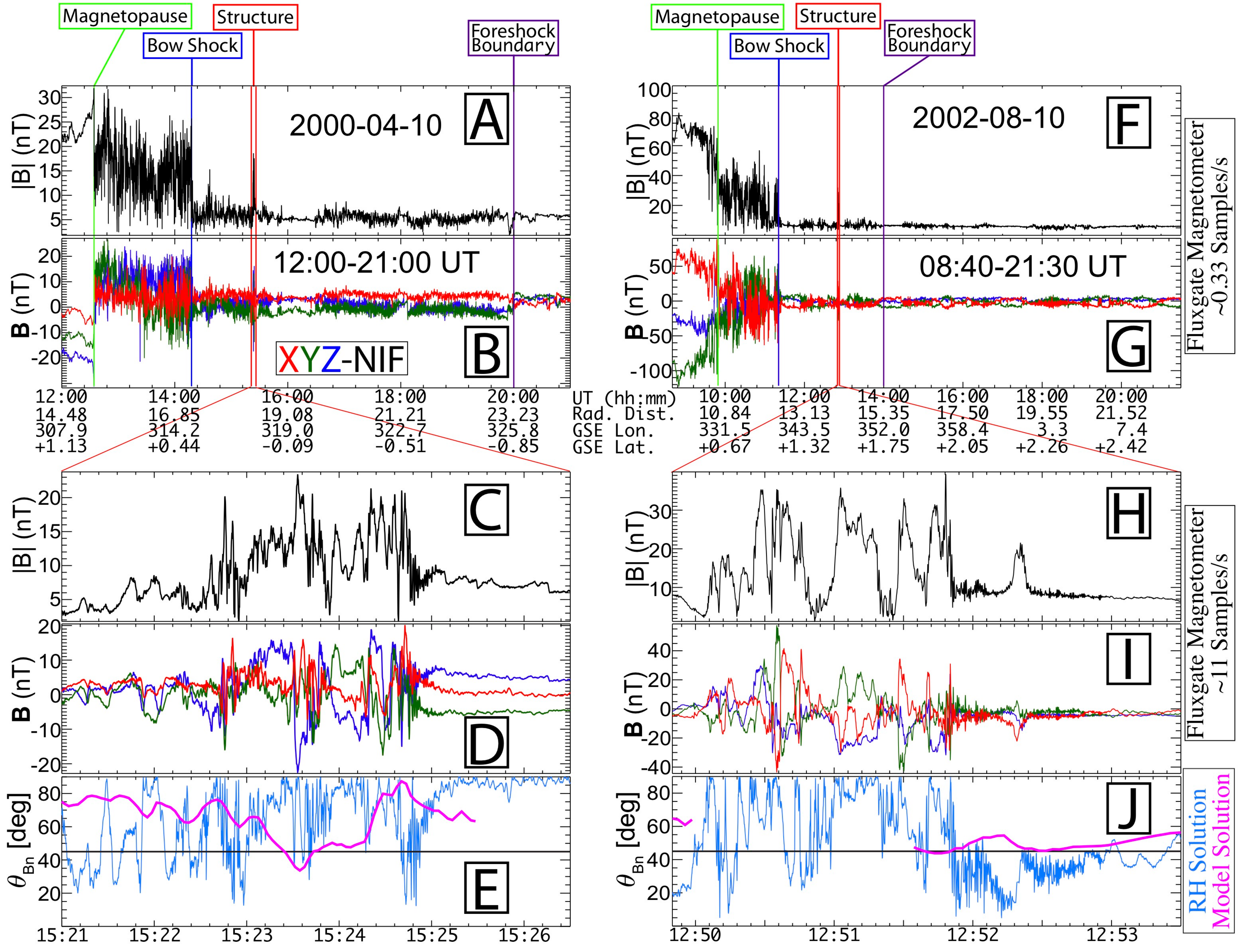}}
    \caption[Overview Plot]{The top half of the figure shows three second resolution of the magnitude and the normal incidence frame (NIF) \citet[\textit{e.g.}][]{sundkvist12a} components of magnetic field data from the Wind spacecraft on 2000-04-10 (\textbf{A}-\textbf{B}) and 2002-08-10 (\textbf{F}-\textbf{G}) each with three vertical lines that indicate the magnetopause crossing (green), the last bow shock crossing (blue), the foreshock boundary (purple), and the red lines show the time periods for panels \textbf{C}-\textbf{E} and \textbf{H}-\textbf{J}.  The tick mark labels at the bottom of these two panels are:  UT time, and the Wind spacecraft radial distance (R${\scriptstyle_{E}}$), GSE longitude (degrees), and GSE latitude (degrees).  Every panel has the same format, but the bottom two panels show the HTR MFI data and the shock normal angle for model (magenta) and Rankine-Hugoniot (blue) solutions.}
    \label{fig:overview}
  \end{center}
\end{figure}

\indent  Figure \ref{fig:overview} plots an overview of the magnetic field measurements observed for the two events of interest, which show structures identified as groups of shocklets and SLAMS.  We will focus on the 2002-08-10 event herein.  First we needed to determine whether the spacecraft was magnetically connected to the terrestrial bowshock and then eliminate the possibility that these structures were simply due to an expansion of or close encounter with the terrestrial bow shock.  Wind was located at a GSE position of $\sim$$<$$+$13.6, -12.4, $+$0.02$>$ R${\scriptstyle_{E}}$ for the group of SLAMS in Figure \ref{fig:overview}\textbf{C}-\textbf{E} and at $\sim$$<$$+$13.8, -3.0, $+$0.4$>$ R${\scriptstyle_{E}}$ for Figure \ref{fig:overview}\textbf{H}-\textbf{J}.  These positions correspond to distances of $\sim$2.4 R${\scriptstyle_{E}}$ and $\sim$1.4 R${\scriptstyle_{E}}$ from the last bow shock crossings, respectively.  \\
\indent  To exclude the possibility of a bow shock expansion, we examined data from the ACE, GOES 8 and 10, and Interball spacecraft (courtesy of CDAWeb).  We found no transient features in any of these data sets suggesting a sudden reduction in solar wind pressure occurred that could allow the bow shock to expand $\gtrsim$1 R${\scriptstyle_{E}}$ out to the Wind position at the time of interest in either event.  Therefore, we conclude that these structures are features of the terrestrial foreshock and not an expansion of the bow shock.  \\
\indent  Next, we needed to determine the local geometry of the bow shock.  The blue line in Figures \ref{fig:overview}\textbf{E} and \ref{fig:overview}\textbf{J} represents the $\theta{\scriptstyle_{Bn}}$ calculated from our first method and the magenta line was calculated using the second method, both discussed in Section \ref{sec:data}.  Note that the second method used a smoothed average magnetic field, instead of the HTR MFI data, and gaps indicate regions not magnetically connected to the model shock surface.  The Rankine-Hugoniot solutions gave us shock normal vectors of $\sim$$<$$+$0.750, -0.621, -0.134$>$ for the 2000-04-10 event and $\sim$$<$$+$0.987, -0.158, $+$0.002$>$ for the 2002-08-10 event.  We estimated the upstream foreshock average magnetic field, \textbf{B}${\scriptstyle_{o}}$, using the Wind/MFI observations between 14:20-15:22 UT for the 2000-04-10 event and between 11:21-12:49 UT for the 2002-08-10 event.  The average GSE vectors were found to be \textbf{B}${\scriptstyle_{o}}$ $\sim$ $\langle$$+$3.60,$-$2.99,$-$1.47$\rangle$ nT and $\sim$ $\langle$$-$4.32,$+$0.65,$-$0.41$\rangle$ nT, respectively.  These estimates with the above shock normal vectors give $\theta{\scriptstyle_{Bn}}$ $\sim$ 14$^{\circ}$ and $\sim$ 6$^{\circ}$, respectively.  Therefore, the Wind spacecraft was primarily traveling through the quasi-parallel region of the terrestrial foreshock for the time of interest.  \\
\indent  The foreshock structures, marked by vertical red lines in Figures \ref{fig:overview}\textbf{A}-\textbf{B} and \ref{fig:overview}\textbf{F}-\textbf{G}, were composed of a series of compressive magnetosonic waves ($\delta$B in phase with density, $\delta$n, fluctuations) identified as shocklets and SLAMS shown in Figures \ref{fig:overview}\textbf{C}-\textbf{D} and \ref{fig:overview}\textbf{H}-\textbf{I}, respectively.  The SLAMS were observed to have:  (1) mixtures of right- and left-hand polarizations (spacecraft frame); (2) very oblique ($\theta{\scriptstyle_{kB}}$ $\gtrsim$ 55$^{\circ}$ and $\theta{\scriptstyle_{kV}}$ $\gtrsim$ 40$^{\circ}$) propagation; and (3) $\delta$B/B${\scriptstyle_{o}}$ $\gtrsim$ 2-6, consistent with previous observations \citep{schwartz92a, wilkinson93a}.  The 2002-08-10 event (Figure \ref{fig:overview}\textbf{H}-\textbf{I}) shows an isolated SLAMS near 12:52:20 UT.  The 2000-04-10 event (Figure \ref{fig:overview}\textbf{C}-\textbf{D}) did not show a similar structure.  The importance of this difference will be discussed in the next section.  \\
\indent  Both groups of SLAMS have higher frequency fluctuations on their leading/upstream (i.e. the right-hand side of Figures \ref{fig:overview}\textbf{C}-\textbf{D} and \ref{fig:overview}\textbf{H}-\textbf{I}) edges, consistent with whistler mode waves.  The characteristics of the waves immediately upstream of the the steepened edges are consistent with previous observations of whistler precursors \citep[\textit{e.g.}][]{wilsoniii09a}.  The whistler amplitudes and beam intensity decrease away from the leading edge of the group of SLAMS and eventually the whistlers disappear when the FABs disappear.  However, we cannot definitively show that the two phenomena are causally related because whistler modes have been observed in the absence of FABs.  While simulations have found that reflected ions can provide free energy for whistler precursors \citep[\textit{e.g.}][]{scholer03b}, supported by recent observations \citep{wilsoniii12d}, their primary source is thought to be dispersive radiation \citep[\textit{e.g.}][]{sundkvist12a}.  The source of the whistler mode waves are beyond the scope of this paper and we will not discuss them further.  
\subsection{Particle Distributions}  \label{subsec:particledists}
\begin{figure}[htb]
  \vspace{-35pt}
  \begin{center}
   {\includegraphics[trim = 0mm 0mm 0mm 0mm, clip, width=14cm]{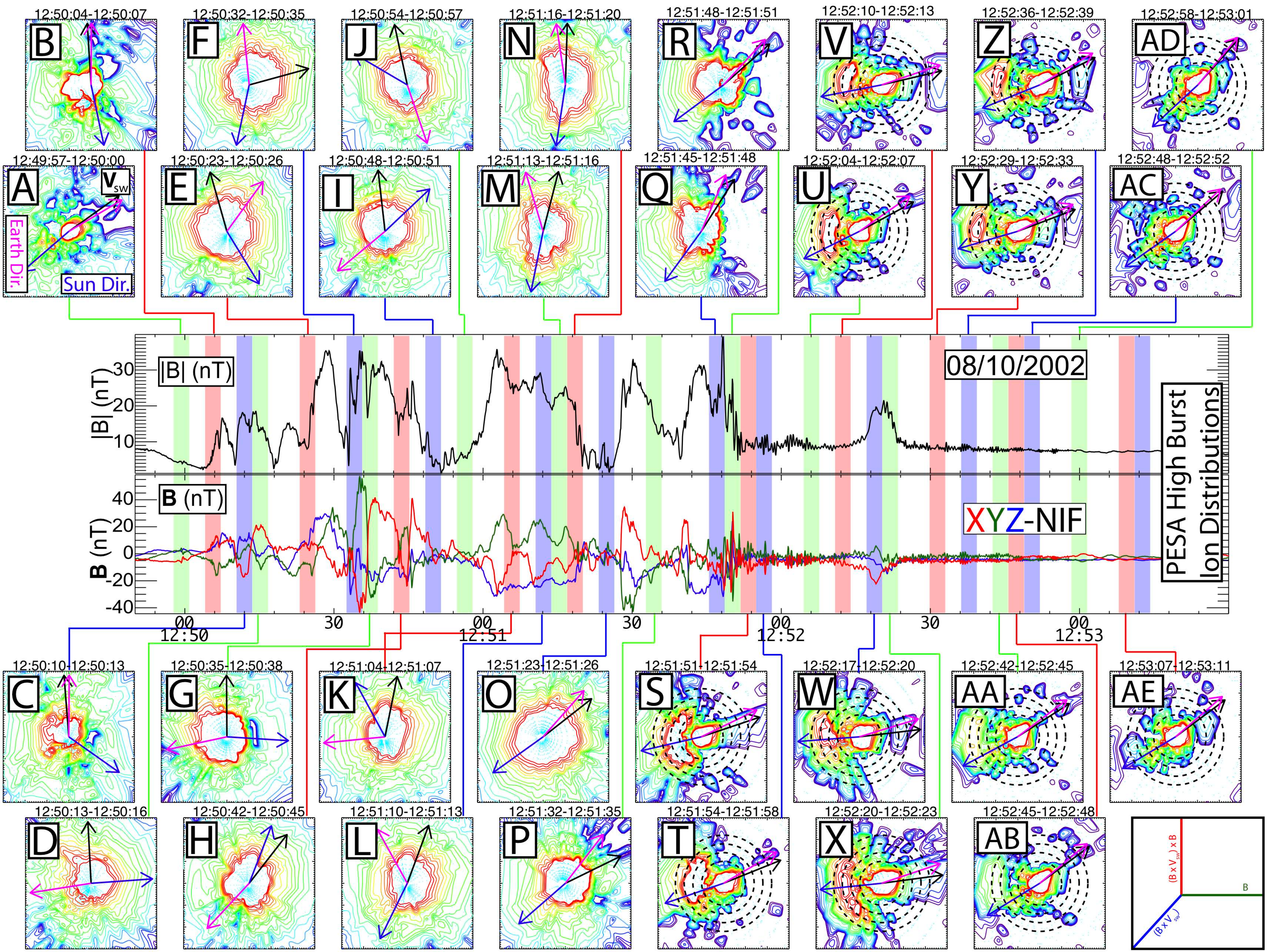}}
    \caption[SLAMS zoom-D to PHB]{Selected PESA High Burst distributions shown for the time range shown in Figure \ref{fig:overview}\textbf{H}-\textbf{J}.  The ion distribution plots are contours of constant phase space density (uniformly scaled from 1$\times$10$^{-13}$ to 1$\times$10$^{-9}$ s$^{3}$cm$^{-3}$km$^{-3}$, where red is high) with arrows showing projections of \textbf{V}${\scriptstyle_{sw}}$ (black), sun direction (blue), and Earth direction (magenta).  Panels \textbf{S}-\textbf{AE} have circles of constant energy at 500, 700, 900, and 1100 km/s.  The coordinate system used for the contours is shown in the lower right-hand corner of the plot, where each plot ranges from $\pm$1500 km/s on each axis.  Ion beams are clearly identified in panels \textbf{B} and \textbf{S}-\textbf{AB}.}
    \label{fig:slamsphb20020810}
  \end{center}
\end{figure}

\begin{figure}[htb]
  \vspace{-35pt}
  \begin{center}
   {\includegraphics[trim = 0mm 0mm 0mm 0mm, clip, width=8.6cm]{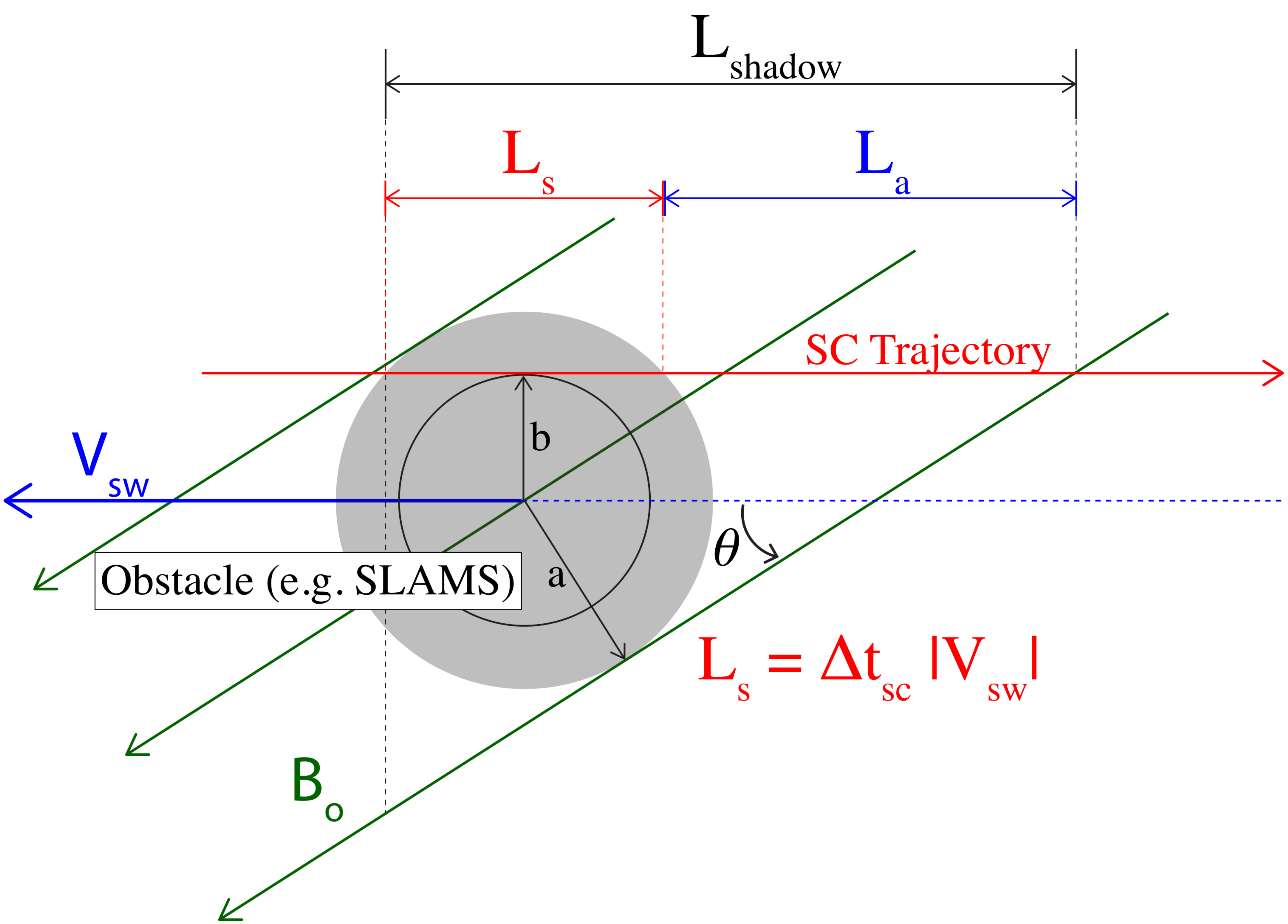}}
     \caption[Example SLAMS Shadow]{A schematic cartoon used to illustrate how the series of SLAMS in Figures \ref{fig:overview}\textbf{C}-\textbf{E} and \ref{fig:overview}\textbf{H}-\textbf{J} can block the spacecraft (SC) from ``seeing'' the bow shock along magnetic field lines.  In this example, the sun is to the right and Earth to the left.}
     \label{fig:cartoon}
  \end{center}
\end{figure}

\indent  We examined the effects caused by the series of shocklets and SLAMS on the ion and electron distribution functions for the two foreshock passes.  The group of SLAMS created a rarefaction region behind the structures with a strong deflection of the solar wind core, analogous to the wake created by an obstacle in a fluid flow.  The SLAMS caused strong anisotropic heating in the low energy ($\lesssim$1.1 keV) electrons and ions ($\lesssim$10 keV).  \\
\indent  Figure \ref{fig:slamsphb20020810} shows the HTR MFI data and select PESA High distribution functions (in the solar wind frame) for the time range corresponding to Figure \ref{fig:overview}\textbf{H}-\textbf{I}.  The solar wind core is clearly identified in the center of panels \textbf{S}-\textbf{AE} and separated from the FABs seen near $\sim$500-900 km/s in panels \textbf{S}-\textbf{AB}, moving anti-parallel to \textbf{B}${\scriptstyle_{o}}$, which corresponds to the sunward direction for this event.  Note that panel \textbf{B} contains a FAB as well, but at lower speeds, which can be seen by the difference between the foreshock observations in panel \textbf{A}.  The FABs on the downstream(earthward) side of the group of SLAMS is weaker for both events.  Hot diffuse ions (i.e. nonthermal tail observed above $\sim$800 km/s) are observed continuously between 12:50:13 UT and 12:51:45 UT, simultaneous with the SLAMS.  The ion core experienced strong enough heating in this region to be observed by PESA High in panels \textbf{D}-\textbf{R} in Figure \ref{fig:slamsphb20020810}, which was supported by the PESA Low and SWE distributions.  These effects are consistent with previous observations \citep{schwartz92a, wilkinson93a}.  \\
\indent  The ion beams shown in Figure \ref{fig:slamsphb20020810} are intense FABs observed between 12:51:51--12:52:48 UT ($\Delta$t${\scriptstyle_{FAB}}$ $\sim$ 57s), which is a considerably longer duration than for the 2000-04-10 event ($\Delta$t${\scriptstyle_{FAB}}$ $\sim$ 10s) but comparable to previous observations \citep[\textit{e.g.}][]{meziane04a}.  The FABs between 12:51:51--12:52:48 UT have T${\scriptstyle_{b}}$ $\sim$ 175-850 eV, T${\scriptstyle_{\perp, b}}$/T${\scriptstyle_{\parallel, b}}$ $\gtrsim$ 2.3-9.7, V${\scriptstyle_{b}}$/V${\scriptstyle_{sw}}$ $\sim$ 1.1-2.4, and n${\scriptstyle_{b}}$/n${\scriptstyle_{i}}$ $\sim$ 0.3-14$\%$, consistent with previous observations \citep[\textit{e.g.}][\textit{and references therein}]{bale05a}.  Correspondingly, the FABs observed at the 2000-04-10 event (not shown) had T${\scriptstyle_{b}}$ $\sim$ 80-200 eV, V${\scriptstyle_{b}}$/V${\scriptstyle_{sw}}$ $\sim$ 1.3-2.0, and n${\scriptstyle_{b}}$/n${\scriptstyle_{i}}$ $\sim$ 0.2-1.6$\%$.  Though we did examine theoretical growth rates associated with ion/ion beam instabilities \citep[\textit{e.g.}][]{akimoto93a}, we do not believe the FABs are the source of the SLAMS, rather we will show that the SLAMS are the source of the FABs.  We believe the source of free energy for SLAMS growth are the diffuse ions observed simultaneously with each group of SLAMS, consistent with theory \citep[\textit{e.g.}][]{scholer03b} and previous observations \citep{schwartz92a}.  \\
\indent  Note that FABs were observed both downstream (earthward) and upstream (sunward) of the group of SLAMS for both events.  The FAB intensity was found to be greater on the upstream(sunward) side of the group of SLAMS but weaker on the upstream side of the isolated SLAMS ($\sim$12:52:20 UT) shown in Figure \ref{fig:overview}\textbf{H}-\textbf{I}.  This suggests that the group of SLAMS may act like a quasi-perpendicular shock, shown in Figures \ref{fig:overview}\textbf{E} and \ref{fig:overview}\textbf{J}, consistent with previous observations \citep{mann94a}.  The decrease in FAB intensity upstream of the isolated SLAMS suggests that groups are necessary to create an obstacle strong enough to reflect a significant population of ions, consistent with theory \citep{schwartz91a}.  If the FABs were produced at the bow shock, then it is very surprising that the beams are so coherent after traversing the turbulent groups of SLAMS.  If the magnetic field cannot connect the spacecraft with the bow shock without going through the SLAMS, then the SLAMS must be the source of the FABs.  \\
\indent  Figure \ref{fig:cartoon} is an illustrative cartoon that we will use to argue that the group of SLAMS in each event is the source of the FABs.  If we assume the group of SLAMS, referred to as the obstacle for brevity, are being convected with the solar wind at roughly \textbf{V}${\scriptstyle_{sw}}$, then the spacecraft will be effectively stationary with respect to the obstacle.  Therefore, the path of the spacecraft through the obstacle is anti-parallel to \textbf{V}${\scriptstyle_{sw}}$ and of length L${\scriptstyle_{s}}$ $=$ V${\scriptstyle_{sw}}$ $\Delta$t${\scriptstyle_{sc}}$.  The spacecraft will be shielded from the terrestrial bow shock along magnetic field lines for a distance L${\scriptstyle_{shadow}}$.  Since we know that \textbf{V}${\scriptstyle_{sw}}$ is at an angle to the background magnetic field, \textbf{B}${\scriptstyle_{o}}$, then L${\scriptstyle_{shadow}}$ $>$($=$) L${\scriptstyle_{s}}$ for angles $<$($=$) 90$^{\circ}$.  It takes $\Delta$t${\scriptstyle_{sc}}$ $\sim$ 136 s(108 s) to traverse the obstacle in the 2000-04-10(2002-08-10) event, much greater than the corresponding $\Delta$t${\scriptstyle_{FAB}}$ discussed above.  The average complementary angle between \textbf{V}${\scriptstyle_{sw}}$ and \textbf{B}${\scriptstyle_{o}}$ is $<$ 45$^{\circ}$ for both events, therefore L${\scriptstyle_{a}}$ $>$ L${\scriptstyle_{s}}$.  This means that the spacecraft is in ``magnetic shadow'' of the obstacle for at least $\Delta$t${\scriptstyle_{sc}}$ after exiting on the upstream side.  Therefore, the spacecraft cannot connect to the bow shock without going through the obstacle.  The amount of turbulence and rotation in the magnetic field observed through the obstacle suggests that a bow shock source would be highly unlikely.  In conclusion, we argue that the group of SLAMS must be the source of the FABs observed on their upstream side, not the bow shock, consistent with predictions of the quasi-parallel bow shock \citep{schwartz91a}.
\section{Discussion and Conclusions}  \label{sec:conclusions}
\indent  This study presents observations of the evolution of field-aligned ion beams (FABs) through SLAMS in the terrestrial foreshock.  The FAB intensities were higher on the upstream(sunward)-side of the group of SLAMS than the downstream.  The group of SLAMS created an effective wall between the spacecraft and the bow shock, which suggests that these structures are locally producing the observed beams.  However, as we discussed the isolated SLAMS near 12:52:20 UT in the 2002-08-10 event may be too thin to locally produce FABs.  In addition, the field is observed to rotate to a quasi-perpendicular orientation within the SLAMS, which would support the previous statement and the predictions of \citet{schwartz91a}.  \\
\indent  The FABs propagate (in the plasma frame) away from the bow shock, consistent with previous observations \citep[\textit{e.g.}][]{kis07a}.  This is also the direction toward the upstream(sunward)-side of the SLAMS.  They had T${\scriptstyle_{b}}$ $\sim$ 80-850 eV, V${\scriptstyle_{b}}$/V${\scriptstyle_{sw}}$ $\sim$ 1.1-2.4, and n${\scriptstyle_{b}}$/n${\scriptstyle_{i}}$ $\sim$ 0.3-14$\%$, consistent with previous observations \citep[\textit{e.g.}][\textit{and references therein}]{bale05a}.  While ion beams have been previously observed near SLAMS and suggested to be locally produced \citep{schwartz92a, wilkinson93a}, no previous reports have shown the evolution of FABs through SLAMS and showed the SLAMS to be the source.  The group of SLAMS was shown to create a barrier between the spacecraft and the bow shock along magnetic field lines, which supports our argument of local beam production and confirms predictions of \citet{schwartz91a}.  \\
\indent  The source of the SLAMS and their associated whistler precursors is beyond the scope of this manuscript, however simulations suggest that diffuse ions are responsible for SLAMS and the combination of dispersion and reflected ion-driven instabilities explain the precursors \citep{scholer03b, sundkvist12a}.  The simultaneous observation of diffuse ions with the SLAMS and FABs with the precursors does support these results, but precursors have been observed in the absence of reflected ions.  \\
\indent  In conclusion, we show the first direct evidence that groups of SLAMS can act like a quasi-perpendicular shock producing reflected field-aligned ion beams.

\section{acknowledgments}  \label{sec:acknowledgments}
  \indent  We thank R. Lin (3DP), K. Ogilvie (SWE), and R. Lepping (MFI) for the use of data from their instruments.  Data from ACE, GOES, Interball, and OMNI data were obtained from CDAWeb.  All data sets from the Wind spacecraft were produced under Wind MO\&DA grants.  This research was supported by NESSF grant NNX07AU72H, grant NNX07AI05G, and the Dr. Leonard Burlaga/Arctowski Medal Fellowship.  \\


\newpage

\end{document}